\documentclass[11pt,a4paper]{article}

\usepackage{graphicx,color}
\usepackage{amsmath}
\usepackage{hyperref}

\usepackage[affil-it]{authblk}
\usepackage[left=2cm,right=2cm,top=2cm,bottom=2cm,a4paper]{geometry}
  
\newcommand{\p}{\partial}
\renewcommand{\d}{\delta}
\newcommand{\av}[1]{\langle  #1 \rangle}

\let\origthanks\thanks
\renewcommand\thanks[1]{\begingroup\let\rlap\relax\origthanks{#1}\endgroup}

\begin{document}

\title{Detector resolution correction for width
of intermediate states in three particle decays}

\author{Igor Denisenko\thanks{iden@jinr.ru}}
\author{Igor Boyko}
\affil{
Joint Institute for Nuclear Research,\\
Joliot-Curie 6, 141980 Dubna, Moscow region, Russia
}

\maketitle

\abstract{We propose a method that allows to take into account detector
resolution in the partial wave analysis event-by-event fit
as a special case.
Implementation of the method is discussed and the applicability of the method is
studied for the $J/\psi \to K^{*\pm}K^{\mp} \to K^+K^-\pi^0$
and $J/\psi \to K_2(1430)^\pm K^{\mp} \to K^+K^-\pi^0$ decays.}

\section{Introduction}
The most of partial wave analyses (PWA) are performed in the framework of the maximum likelihood method.
For a typical set-up the log-likelihood function is $s=-\sum_i\ln P_i$, where $i$ runs over selected data events
and $P_i$ is the probability to observe an event with measured momenta of final particles $p_i$. The latter
is given by $P_i = \epsilon_i\sigma_i / \int (\epsilon\sigma)d\Phi$, where $\epsilon_i$
is the selection efficiency for the kinematics of event $i$, $\sigma_i=\sigma(p_i)$ is
the differential cross section (depending on the fitted parameters) and the integral gives the overall
normalization factor. To take into account the detector resolution one has to introduce
a convolution with detector response function: $\epsilon\sigma \to R\otimes (\epsilon\sigma)$.
In general case such convolution can not be performed within reasonable CPU time. Here
we show that for a special case discussed below the detector resolution can be taken into
account. 

Without loss of generality the method will be described in application
to the $J/\psi \to K^{*\pm}K^{\mp} \to K^+K^-\pi^0$ decay, where we are
interested in measuring the width of $K^*$ and assume that $J/\psi$
is produced in a $e^+e^-$ collider experiment. We also apply the
method to $J/\psi \to K_2(1430)^\pm K^{\mp} \to K^+K^-\pi^0$.

\section{Method}
Typically the cross section is calculated from measured momenta of
final state particles. For the following we consider the case when these momenta are
taken after the kinematic fit, so that the total four-momentum
is known precisely and is not smeared.

The method is based on three main assumptions:
\begin{itemize}
    \item
        We assume that the kinematical variables the cross section depends on
        can be divided into two groups such that in the calculation of the
        convolution of cross section
        with the detector response function, the cross section dependence on the
        second group of variables can be neglected. The qualitative definition
        will be given later.
    \item The mass resolution in the studied kinematic channel is much smaller
        than the width of the studied resonance (in our example
        $\sigma_{K\pi} \ll \Gamma_{K^*}$, where $\sigma_{K\pi}$ stands for the
        mass resolution in the $K\pi$ channel).
    \item Monte-Carlo simulation of detector resolution and efficiency
        is consistent with real data processing.
\end{itemize}
From general arguments the differential cross section of the $J/\psi\to K^+K^-\pi^0$ decay
in the center-of-mass reference frame depends on four variables
($J/\psi$ in relativistic $e^+e^-$ collisions is produced with $J_z=\pm1$,
so we measure $3\times3$ particle momenta with overall 4-momentum constraint
and rotation symmetry along the beam axis)
denoted by $a_k$, $k=1..4$.
The Breit-Wigner parts of the amplitude 
depend on invariant masses of any two pairs of final particles
(the invariant mass of the third pair depends on the first two), for computation
convenience we took $M^2(K^+\pi^0)$ and $M^2(K^-\pi^0)$. The other two variables
are arguments of the angular parts of the amplitude and we do not specify them.

For an event with the measured momenta $q$, the convolution reads:
\begin{equation}
\begin{split}
    \nonumber
  \label{pi_i_rc}
(R\otimes\sigma\epsilon)(q) &= \int R(q'-q,q) \epsilon(q') \sigma(a(q')) dq' \approx \\
                           &\approx \epsilon(q) \int R(q'-q,q) \sigma(a(q')) dq', \\
\end{split}
\end{equation}
where $R(q'-q,q)$ is the detector response function (note, due to the kinematic fit
not all momenta here are independent and that is taken into account in $R(q'-q,q)$).
For the following we will not need the explicit form of $R(q'-q,q)$. Then we take integrals and
change variables to $a$:

\begin{equation}
\begin{split}
  \label{eq:taylorexp}
\int R(q'-q,q) \sigma(a(q')) dq' =  \int \tilde R(a'-a^q,q) \sigma(a') da' = \\
=\int \tilde R(\d a, q) \biggl( \sigma(a^q) + 
\sum_k\frac{\p \sigma(a)}{\p a_k}\d a_k + \sum_{k,l}\frac{1}{2}\frac{\p^2 \sigma(a)}{\p a_k \p a_l} \d a_k \d a_l + \\
  + O(\d a^3) \biggr) d\d a 
= \sigma \Big|_{a^q} + \sum_k\frac{\p \sigma}{\p a_k} \Big|_{a^q} \cdot \av{\d a_k} + \\
+ \sum_{k,l}\frac{1}{2}\frac{\p^2 \sigma}{\p a_k \p a_l}\Big|_{a^q} \cdot \av{\d a_k \d a_l}
   + \int \tilde R(\d a, q) O(\d a^3) d\d a.
\end{split}
\end{equation}
Here averages $\av{\d a_k}$ and $\av{\d a_k \d a_l}$ are calculated using 
$\tilde R(\d a, q)$ as weight. The last step is
to take into account in equation~\ref{eq:taylorexp} only variables that cross section
strongly depends on. Quantitatively we assume that
\begin{equation}
    \nonumber
\begin{split}
   \sum_\alpha \frac{\p \sigma}{\p a_\alpha} \Big|_{a^q} \cdot \av{\d a_\alpha} 
   + \sum_{\alpha,\beta}\frac{1}{2}\frac{\p^2 \sigma}{\p a_\alpha \p a_\beta}\Big|_{a^q} \cdot \av{\d a_\alpha \d a_\beta} 
   \gg \\
   \sum_\zeta\frac{\p \sigma}{\p a_\zeta} \Big|_{a^q} \cdot \av{\d a_\zeta} 
   + \sum_{\zeta\xi}\frac{1}{2}\frac{\p^2 \sigma}{\p a_\zeta \p a_\xi}\Big|_{a^q} \cdot \av{\d a_\zeta \d a_\xi},
\end{split}
\end{equation}
where indexes $\alpha$ and $\beta$ run over the first group of variables 
($M^2(K^+\pi^0)$ and $M^2(K^-\pi^0)$), $\zeta$ and $\xi$ run over
the second group. Finally one gets the computation formula:
\begin{equation}
\label{eq:corrused}
\begin{split}
       (R\otimes\sigma)(q) \approx\\
       \sigma(q)  + \sum_\alpha\frac{\p \sigma}{\p M^2_\alpha} \Big|_{(q)}
          \av{\d M^2_\alpha} +
       \sum_{\alpha,\beta} \frac{1}{2}\frac{\p^2 \sigma}{\p M^2_\alpha \p M^2_\beta}
       \Big|_{(q)} \av{\d M^2_\alpha \d M^2_\beta}.
     \end{split}
\end{equation}
Derivatives and averages are calculated numerically,
the averages can be taken as constants in the fit.  
Despite of the chosen variables, this expression can be also
used for resonances in the $K^+K^-$ channel.

\section{Notes on implementation}
The decay amplitude in a particular kinematic channel is a product
of angular and Breit-Wigner-like parts. So, calculating cross section
derivatives in the given approximation, one can vary only invariant mass
squared keeping the angular part constant, which can be easily implemented
in the existing PWA code. For our tests the first and the second cross section
derivatives were calculated in the simplest finite-difference scheme (it required
calculation of the Breit-Wigner amplitude part in six additional points).

As resolution is much smaller than the width of the studied intermediate
resonance the fit can be done  iteratively: the first fit
is performed ignoring detector resolution effects, averages in equation
equation~\ref{eq:corrused} for the obtained solution
are calculated, than the fit that takes into account the detector resolution is performed.
For the cases considered below we saw no improvement of the final results
if an additional iteration was performed.

\section{Method performance: numerical study}
\label{sec:numerical}
In our numerical study we assume a typical set-up of
a $e^+e^-$ collider experiment (e.g. see \cite{Ablikim:2009aa}).
We model the detector response by smearing
generated particle momenta.
To get an estimate of the method sensitivity to particular
shape of the detector response, we consider two
detector response models (referred in the following
as ``model 1'' and ``model 2'').
In the first model we add Gaussian fluctuations
with zero mean to track helix parameters
($\kappa$, $\tan\lambda$, $\phi_0$).
For photons the energy and the direction are smeared ($e^{ph}$, $\theta^{ph}$, $\varphi^{ph}$)
in the same way.
To study the method applicability for different $K^\pm\pi^0$ invariant mass
resolution we use set generated samples with variances proportional to
\begin{align}
\sigma_\kappa/\kappa    &= 5\times10^{-3},  \nonumber \\
\sigma_{\tan\lambda}    &= 5\times10^{-3},  \nonumber \\
\sigma_{\varphi_0}      &= 2.5\times10^{-3} \nonumber \\
\nonumber
\end{align}
for kaon tracks and
\begin{align}
\sigma_{e^{ph}}/e^{ph}   &=2.5\times10^{-3},\nonumber \\
\sigma_{\theta^{ph}}     &=1\times10^{-2},  \nonumber \\
\sigma_{\varphi^{ph}}    &=1\times10^{-2}   \nonumber \\
\nonumber
\end{align}
for photons. As we mentioned above, the input for PWA are particle
four-momenta after the kinematic fit (here the fit is additionally constrained
for the $\pi^0$ mass).

In the second model we vary Cartesian momentum projection of $K^+$, $K^-$, $\pi^0$
with standard deviations proportional to
\begin{align}
    \sigma_{p^K_x} = \sigma_{p^K_y} = \sigma_{p^K_z}&= 3\text{ MeV},  \nonumber \\
    \sigma_{p^{\pi^0}_x} = \sigma_{p^{\pi^0}_x} = \sigma_{p^{\pi^0}_z}  &= 5\text{ MeV}.  \nonumber \\
\nonumber
\end{align}

Reactions $J/\psi \to K^*(892)^\pm K^\mp \to K^+K^-\pi^0$ and
$J/\psi \to K_2(1430)^\pm K^\mp \to K^+K^-\pi^0$ are modeled by weighting
a phase space distributed sample. For each set of smearing parameters
the fitting procedure is repeated several times with different
generated phase space samples of approximately
$0.9\times10^6$ events.

\begin{figure}[h]
    \begin{center}
        \includegraphics[width=.45\textwidth]{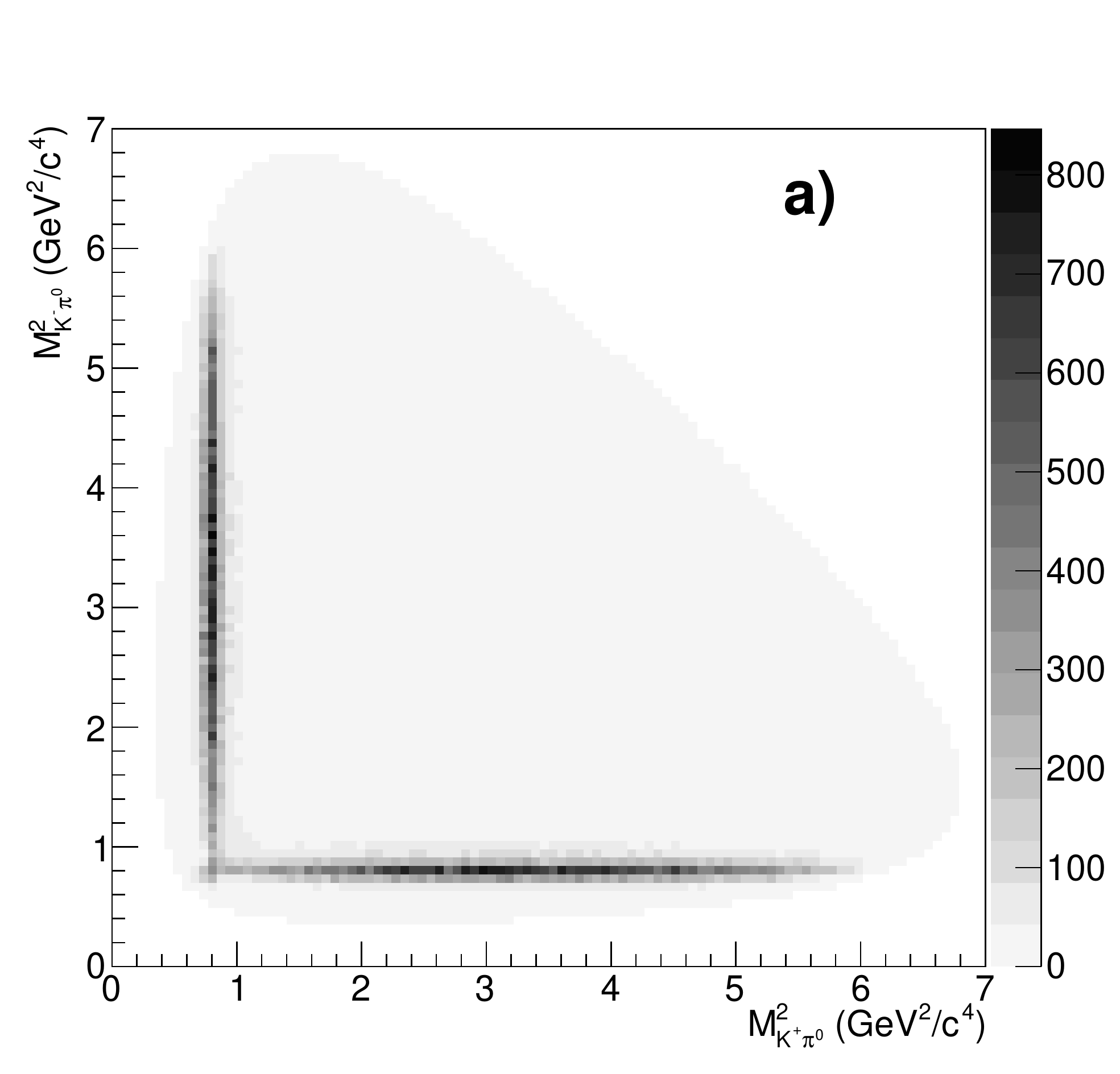}
        \includegraphics[width=.45\textwidth]{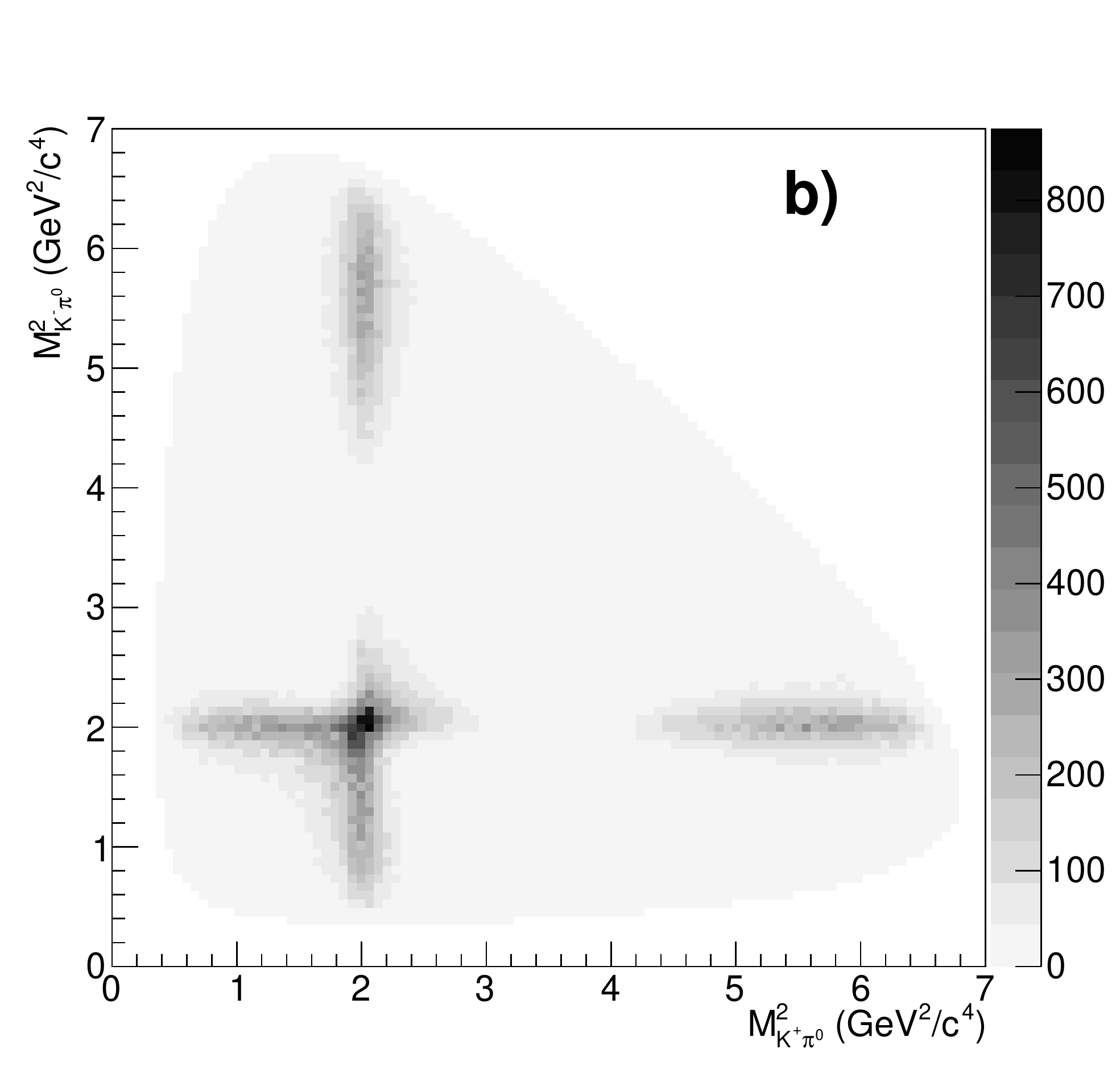}
    \end{center}
    \caption{a)-b) Dalitz plots for generated $J/\psi\to K^*(892)^\pm K^\mp\to K^+K^-\pi^0$  and $J/\psi\to K_2(1430)^\mp K^\mp\to K^+K^-\pi^0$ samples correspondingly.}
    \label{fig:dalitz-input}
\end{figure}

The decay is parametrized in the covariant
tensor formalism framework \cite{Anisovich:2004zz}, for the Breit-Wigner part we 
used
\begin{equation}
\nonumber 
\begin{split}
A^{BW}       &= \frac{1}{M^2-s-iM\Gamma_{J}(s)},\\
\Gamma_{J}(s)&=\frac{\rho(s)}{\rho(M^2_{J})}\Gamma,\\
\rho(s)      &=\frac{2k}{\sqrt{s}}\frac{k^{2J}}{F^2(k^2,r_{J}^2,J)}.
\end{split}
\end{equation}
Here $s$ and $k$ are the invariant mass squared and relative momentum of the
resonance daughter particles, $J$ is the spin of the resonance; $M$, $\Gamma$
and $r$ are mass, width and Blatt-Weisskopf radius correspondingly; functions
$F(k^2,r_{J}^2,J)$ are Blatt-Weisskopf form factors, which can be found in \cite{Anisovich:2004zz}.
This later is the most suitable parametrization for $K^*(892)$ and can be also
applied for $K_2(1430)$. In our study we use PDG averages \cite{Agashe:2014kda} for
the mass and the width
of $K^*(892)$ and $K_2(1430)$ and fix their Blatt-Weisskopf radii to $0.5$ fm.
Dalitz plots for the ``generated data samples'' are shown in figure~\ref{fig:dalitz-input}.

Averages $\av{\d a_\alpha}$ and $\av{\d a_\alpha \d a_\beta}$
are determined iteratively as explained above.

\begin{figure*}[h]
    \begin{center}
        \includegraphics[width=.80\textwidth]{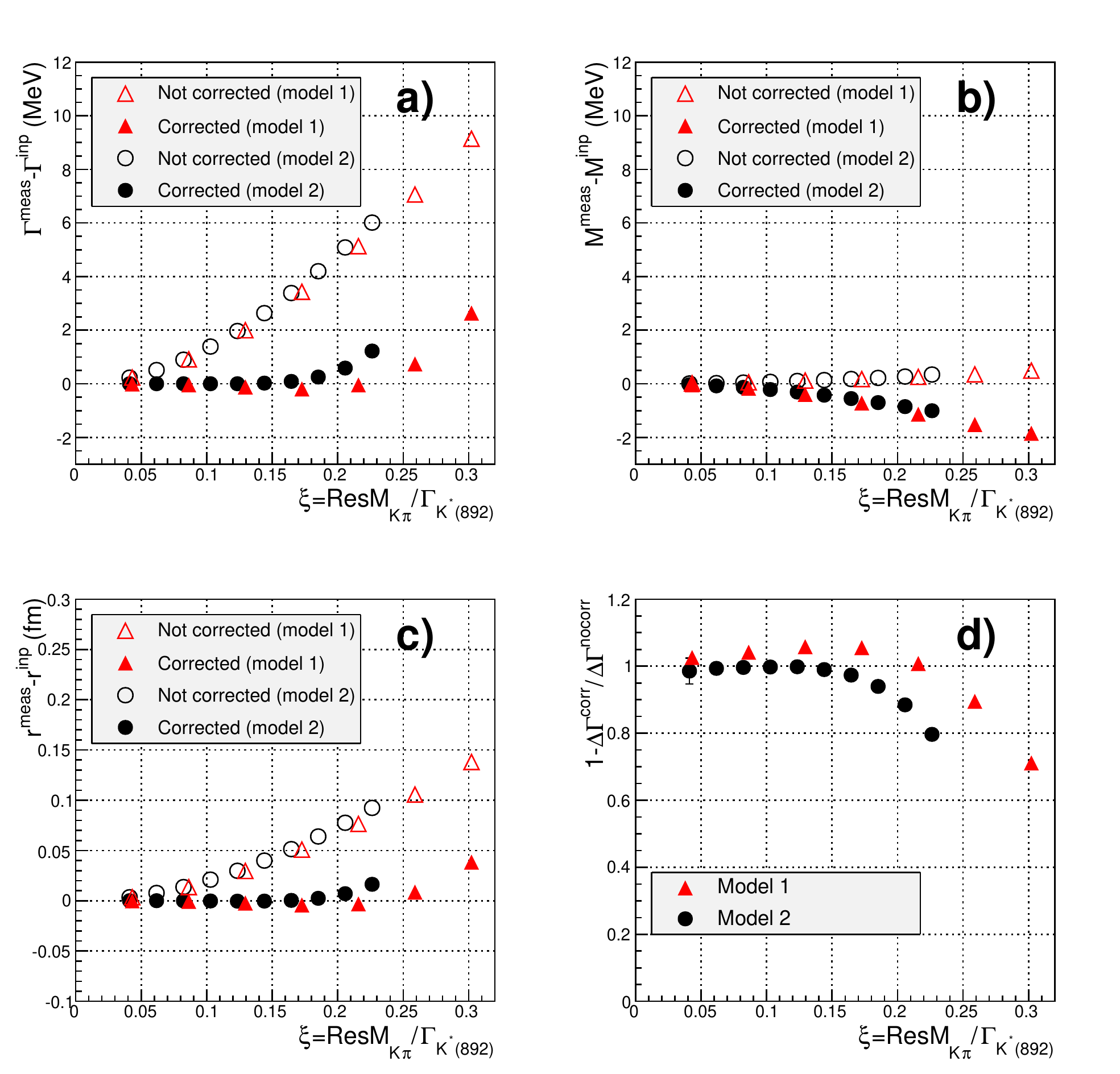}
    \end{center}
    \caption{A comparison of fit results when detector resolution is
    taken into account (solid markers) and ignored (open markers)
    for $K^*(892)$. Results for the first and the second (see in the text)
    detector response models are shown in red and black correspondingly.
    Figures a)-c) show fitted width, mass and Blatt-Weisskopf radius,
    figure d) shows the fraction of width bias due to detector response
    compensated by the used method. The shown results for the model 2
    are limited by $\xi<0.23$ as for higher $\xi$ values using of the
    quadratic approximation results in non-positive differential
    cross section for some ``data points''. }
    \label{fig:k892-perf}
\end{figure*}

The comparison of fit results that ignore and take into account
the correction for the detector resolution are shown in figure~\ref{fig:k892-perf}.
Deviation of mass, width, Blatt-Weisskopf radius and the
fraction of corrected width bias are given
as a function of the $\xi = Res M_{K\pi}/\Gamma$ ratio,
where $Res M_{K\pi}$ is the variance of the $M_{K\pi}$.
The invariant mass variance is taken
in the resonance band region (i.e. for $K^*(892)$
it is calculated in the band 
$0.8\text{ GeV} < M_{K^+\pi^0} < 1.0\text{ GeV}$ and
$M(K^-\pi^0) > 1.0$ GeV).
The results are provided for two detector response models.

\section{Discussion}
Firstly, we see that the used method allows
to compensate approximately 80-90\% of the resonance width 
and Blatt-Weisskopf radius biases due to the detector
resolution for $\xi < 0.2$. The method applicability
dramatically decreases at higher $\xi$ values.

Secondly, due to made approximations the applied method
introduces bias to the fitted mass of the resonance.
In the case of $K^*(892)$ it is almost independent
of the detector response model and equals to
$0.2$~MeV for $\xi=0.1$.
If we increase the mass of the resonance 
this bias decreases and finally changes sign.
In the case of $K_2(1430)$
taking into account detector resolution can
improve or worsen fit results depending on
the detector response model. 
However, sometimes event reconstruction itself can bias measured 
mass of two particles
(for example due to non-Gaussian detector response to photons).
In the proposed method such effects are taken into account in the linear term in
equation~\ref{eq:corrused} and possibly can improve overall mass measurement.

Thirdly, we also apply this method to resonances in the $K^+K^-$
kinematic channel and find results similar to the presented.

We found the fit results not essentially dependent on used
detector response model, but some difference tells us
that MC study is needed in any practical
application of the method.

The computation time increased approximately proportional to the number of
additional points used to calculate the cross section derivatives.

\begin{figure*}[h]
    \begin{center}
        \includegraphics[width=.80\textwidth]{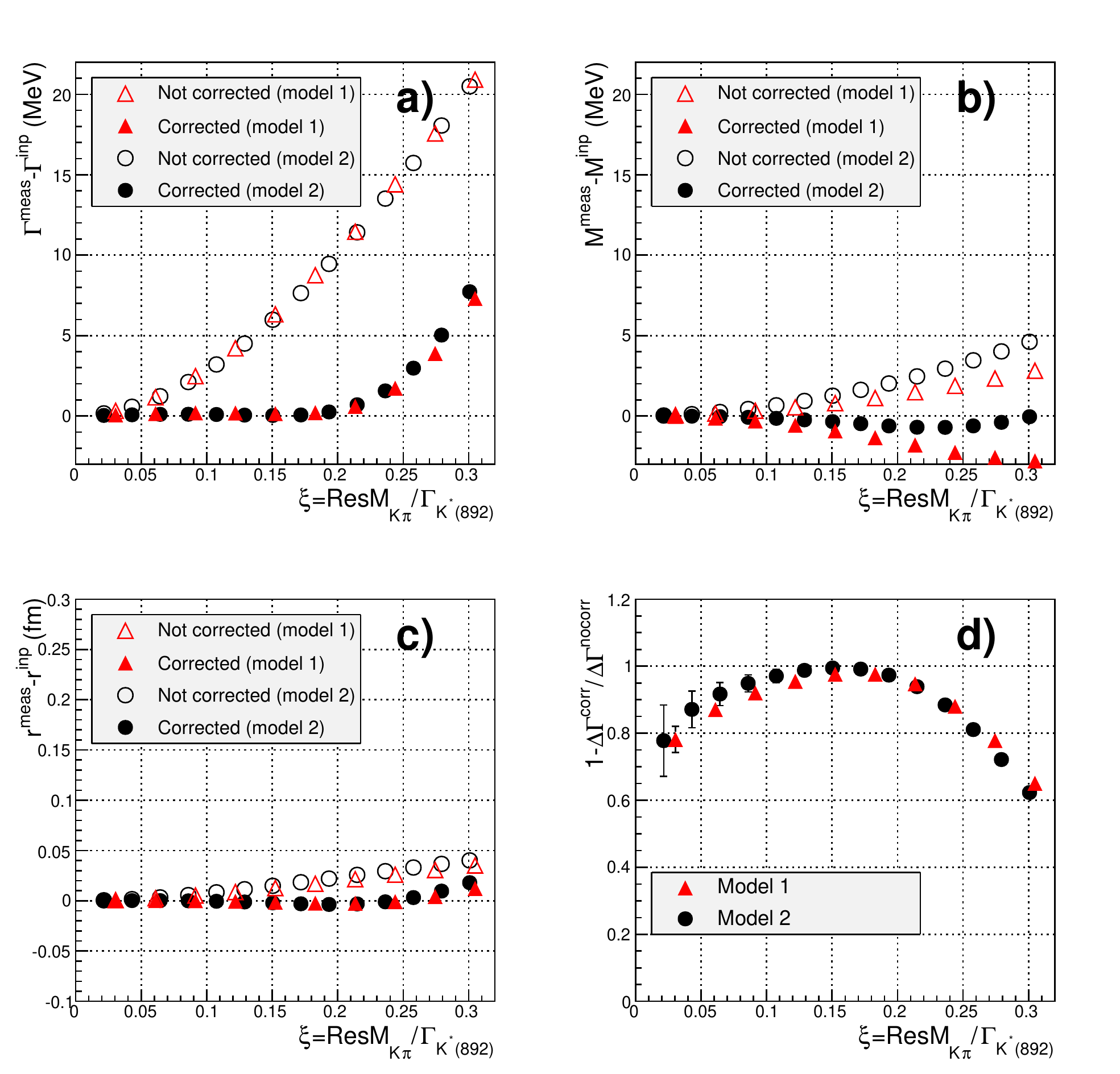}
    \end{center}
    \caption{A comparison of fit results when detector resolution is
    taken into account (solid markers) and ignored (open markers)
    for $K_2(1430)$. Results for the first and the second (see in the text)
    detector response models are shown in red and black correspondingly.
    Figures a)-c) show fitted width, mass and Blatt-Weisskopf radius,
    figure d) shows the fraction of width bias due to detector response
    compensated by the used method.}
    \label{fig:k1430-perf}
\end{figure*}

\section{Conclusion}

For the first time we propose a practically acceptable method,
which allows in special cases to take into account the detector resolution
in the event-by-event partial wave analysis fit.
The method reduces the computation of convolution of a process cross
section and a detector response function to calculating cross section
derivatives and tabulating variances of essential cross section arguments.
We demonstrate the method performance and applicability in the set-up
of a typical $e^+e^-$ experiment for two toy detector response models
and two reactions: $J/\psi \to K^*(892)^\pm K^\mp \to K^+K^-\pi^0 $ and 
$J/\psi \to K_2(1430)^\pm K^\mp \to K^+K^-\pi^0$.


\begin{thebibliography}{9}
\bibitem{Ablikim:2009aa}
 M.~Ablikim {\it et al.},
  Nucl.\ Instrum.\ Meth.\ A {\bf 614} (2010) 345.

\bibitem{Anisovich:2004zz}
  A.~Anisovich, E.~Klempt, A.~Sarantsev and U.~Thoma,
  Eur.\ Phys.\ J.\ A {\bf 24} (2005) 111.


\bibitem{Agashe:2014kda}
  K.~A.~Olive {\it et al.},
  Chin.\ Phys.\ C {\bf 38} (2014) 090001.

\end{thebibliography}
\end{document}